\documentclass[11pt]{article}
\usepackage[latin1]{inputenc}
\usepackage[english]{babel}
\usepackage[namelimits]{amsmath}
\usepackage{authblk}
\usepackage{amssymb}
\usepackage{amsmath}
\usepackage{amsthm}

{\tiny}

\begin{document}
\title{{\bf The physical nature of the cosmological constant and the decoherence scale in a renormalization-group approach}}
\author[1,2,3]{S. Viaggiu\thanks{s.viaggiu@unimarconi.it and viaggiu@axp.mat.uniroma2.it}}
\affil[1]{Dipartimento di Scienze Ingegneristiche, Universit\'a degli Studi Guglielmo Marconi, Via Plinio 44, I-00193 Roma, Italy.}
\affil[2]{Dipartimento di Matematica, Universit\`a di Roma ``Tor Vergata'', Via della Ricerca Scientifica, 1, I-00133 Roma, Italy.}
\affil[3]{INFN, Sezione di Napoli, Complesso Universitario di Monte S. Angelo, Via Cintia Edificio 6, 80126 Napoli, Italy.}

\date{\today}\maketitle
\begin{abstract}
\noindent In this paper we consider the nature of the cosmological constant as due by quantum fluctuations. Quantum fluctuations are generated at Planckian scales by noncommutative effects and watered down at larger scales up to a decoherence scale $L_D$ where classicality is reached.
In particular, we formally depict the presence of the scale at $L_D$ by adopting a renormalization group approach. 
As a result, an analogy arises between the expression for the observed cosmological constant 
$\overline{\Lambda}$ generated  by quantum fluctuations and the one expected by a renormalization group approach, 
provided that the renormalization scale $\mu$ is suitably chosen.
In this framework, the decoherence scale $L_D$ is naturally identified with the value ${\mu}_D$, with  $\hbar{\mu}_D$
representing the minimum allowed particle-momentum 
for our visible universe. 
Finally, by mimicking renormalization group approach, we present a technique to formally obtain a non-trivial infrared (IR) fixed point at $\mu=\mu_D$ in our model.
\end{abstract}

PACS Number(s): 95.36.+x, 98.80.-k, 04.60.Bc, 05.20.-y, 04.60.-m
\section{Introduction}

As is well known, about $68\%$ of the universe is composed of dark energy depicted, in the concordance $\Lambda$CDM model, by a cosmological constant. The cosmological constant is provided by a constant (both in time and space) energy density given by ${\rho}_{\overline{\Lambda}}=\frac{ c^2}{8\overline{\Lambda}\pi G}$, with the exotic equation of state $p_{\overline{\Lambda}}=-c^2{\rho}_{\overline{\Lambda}}$, with $p_{\overline{\Lambda}}$ the pressure. Unfortunately, the nature and origin of $\overline{\Lambda}$ is still matter of debat (see for example \cite{4,5,6,7,8,9,10,11,12,13,14} and references therein) and a complete physically viable description of the cosmological constant is still
lacking. The equation of state for $\overline{\Lambda}$ strongly suggests its relation with vacuum fluctuations. Unfortunately, after introducing a Planckian ultraviolet cutoff $L_P$ and adopting simplistic arguments of quantum field theory, one has
${\overline{\Lambda}}\sim\frac{1}{L_P^2}$, leading to a value of about $122$ order greater than the one effectively observed. This is the 'Planckian catastrophe', the wrongest prediction in the hystory of physics, as stated by Leonard Susskind. To this purpose, in \cite{a1} it has been shown that by using renormalization
techniques with the help of the adiabatic regularization procedure, the renormalized vacuum energy density does not contain terms proportional to the quartic power of the mass, 
thus alleviating the Planckian catastrophe issue.
In \cite{11} 
quantum fluctuations have been depicted in terms of a stochastic inhomogeneous metric. The model is plagued by several issues (regularization used is not Lorentz-invariant) and has been corrected in \cite{12}, but with a problematic assumption of a micro-cyclic universe with an exotic 'micro'
big bang singularity. In \cite{13} a dynamical dark energy in terms of  Ashtekar variables is studied. In \cite{14} the cosmological constant is supposed to be effectively given by ${\overline{\Lambda}}\sim\frac{1}{L_P^2}$ at Planckian scales, but effective  inhomogeneities at Planckian scales inhibit a huge cosmological constant and as a consequence a small cosmological constant emerges at macroscopic scales. In \cite{1,2,3}, we presented a new model. The new idea is that a huge cosmological constant is effectively created at Planckian scales, where Planckian fluctuations are expected to be very strong, but vacuum fluctuations must be averaged on bigger and bigger scales. Hence, with respect to the work in \cite{14}, 
the cosmological constant is not inhibited by inhomogeneities at Planckian scales, but rather quantum fluctuations have an effective monotonic 
decreasing dependence on the physical scale under consideration. 
As a result, an effective cosmological constant emerges it depending on the physical scale under consideration up to a scale $L_D$, denoted as decoherence scale, 
where the crossover to classicality is obtained. The scale $L_D$ fixes the observed value of $\overline{\Lambda}$. In this paper we continue the investigations in \cite{1,2,3}. In particular, after adopting a suitable physically reasonable expression for the renormalization scale $\mu$,
we analyse the possible relation between our semiclassical expression for ${\rho}_{\overline{\Lambda}}$ 
and the one extrapolated in literature by using a renormalization group approach \cite{15a,17,17a,15,16,16a}.\\
In section 2 we present the main features of our model, and in section 3 a renormalization group approach in a cosmological context is summarized. 
In section 4 the parameter $\mu$ is fixed together with its relation with the decoherence scale $L_D$. In section 5 we build a non trivial fixed point at $\mu=\mu_D$.
Finally, section 6 is devoted to some conclusions and final remarks.

\section{Cosmological constant semiclassical model and the decoherence scale}

The classical metric of a de Sitter universe with zero spatial curvature in comoving coordinates is given by
\begin{equation}
ds^2=-c^2dt^2+a^2(t)\left(dr^2+r^2d\theta^2+r^2\sin^2\theta\;d\phi^2\right),
\label{1}	
\end{equation}
where $a(t)\sim e^{ct\sqrt{\frac{\overline{\Lambda}}{3}}}$. With $\overline{\Lambda}$ we denote the actual observed value of the cosmological constant.
First of all note that \cite{1,2,3}, thanks to homogeneity and isotropy of the spacetime, we can consider an arbitrary spherical region of the universe centered in an arbitrary spatial point. At Planckian scales this spherical ball will result inhomogeneous.
This inhomogeneous metric is averaged by using the Buchert scheme \cite{21} but at miscroscopic scales. With the template homogeneous and isotropic metric so obtained
we can introduce a noncommutative spacetime \cite{18,19} where the results in \cite{20} suitable in a Friedmann flat background can be adopted. In particular, it is possible to obtain a semiclassical expression for quantum fluctuations within an arbitrary ball of proper areal radius $L$. At the decoherence scale $L_D$ classicality is
reached and the classical metric (\ref{1}) emerges with the observed value $\overline{\Lambda}$ of the cosmological constant. As a consequence, any spherical region with
$L\geq L_D$ will expand with the same factor $a(t)$ in (\ref{1}), as implied by homogeneity and isotropy.
To be more quantitative, the starting point of the model presented in \cite{1,2,3} is to consider a spherical ball of fixed proper areal radius $L=a(t)r$. For the aforementioned spherical ball, general relativity allows to define a quasi-local 
(Misner-Sharp) mass $M_{ms}$, with related Misner Sharp energy given by
$E_{ms}=M_{ms}c^2=\frac{c^4}{2G}\frac{L^3}{L_{A}^2}$, with $L_A^2=\frac{3}{\overline{\Lambda}}$. Misner-Sharp mass represents, in a cosmological background,
the total energy enclosed inside a spherical shell of proper areal radius $L$ of the cosmic fluid composing the spacetime.
In the Newtonian limit, the Misner-Sharp mass reproduces the rest mass of a spherical body together with its
potential energy. Hence, the Misner-Sharp mass is a good choice to describe the energy content of a spherical configuration in a 
dynamic cosmological background, where the metric is no longer asymptotically flat.
In order to explain the existence of a fluid equipped with the equation of state suitable for $\overline{\Lambda}$ (or more generally a $\gamma-$linear equation of state
$c^2\rho=\gamma p$), a physical mechanism is proposed in \cite{1}: 
quantum fluctuations are capable to transform a radiation field into a one with the equation of state of $\overline{\Lambda}$ at Planckian scales or soon above. To this regard, a theorem is proved in \cite{1,2,3} giving the suitable quasi-local energy $E$ in the form
\begin{equation}
E=E^{(0)}+\hbar\;\Phi(L).
\label{2}
\end{equation}
where $E^{(0)}$ is the energy of an initial radiation field and $\Phi(L)$ is a function satisfying the ordinary differential equation:
\begin{equation}
\hbar\left[L\;{\Phi}_{,L}(L)+\Phi(L)\right]=E(L)(1-3\gamma), \;\;E(L)-\hbar\;\Phi(L) > 0\label{3}
\end{equation}
with $\gamma=-1$ for $\overline{\Lambda}$ and commas denotes partial derivative. As shown in \cite{1,2,3}, after using for $E(L)$ the classical expression $E_{ms}=\frac{c^4}{2G}\frac{L^3}{L_A^2}$ provided by the Misner-Sharp energy, we obtain the case with $\gamma=-1$ only in the 'bare'
case.\footnote{In \cite{1,2,3} the usual decomposition for the observed cosmological constant given by 
$\overline{\Lambda}=\Lambda+{\Lambda}_{vac}$ is adopted, where $\Lambda$ is the bare cosmological constant dressed by the term
${\Lambda}_{vac}$ due to quantum fluctuations}
To obtain the 'dressed' case, we need the generalization of the classical Misner-Sharp energy due to  
general relativity by means of a quantum treatment of Planckian fluctuations. Unfortunately,
a complete quantum gravity theory is not currently at our disposal. As a consequence, in \cite{1,2,3} a semiclassical model is presented by means of a non-commutative spacetime \cite{18,19,20} at Planckian scales or above. By the term 'semiclassical' we mean that metric is considered classical but coordinates become quantum 
non-commutating operators.
To this purpose, the metric at such scales is depicted in our semiclassical model as a inhomogeneous one \cite{2} written in the usual form
\begin{equation}
ds^2=-N(t, x^i)c^2 dt^2+h_{ij}(t, x^i)dx^i dx^j,
\label{4}
\end{equation}
where coordinates $\{x^{\mu}\}$ in (\ref{4}) can be seen as mean values of the coordinates operator
with respect to a state with spherical symmetry in a noncommutative spacetime (see \cite{1,2}).
In order to deal with this scenario, in \cite{1,2,3} the fitting problem present in cosmology is translated at miscroscopic scales by means of a suitable modification of Buchert method \cite{21}. To this averaged template spacetime so obtained, we can apply the reasonings in \cite{20} and thus calculate a physically motivated expression 
for the generalized Misner-Sharp mass corrected by quantum fluctuations.
Spacetime uncertainty relations \cite{20} (STUR) can be deduced in terms of the time coordinate $t$ and proper coordinates
${\eta}^i= a(t) x^i$ with $x^i$ Cartesian coordinates. The next step is to consider maximal localizing states $\{s\}$\cite{18}, that are spherical states 
where coordinate uncertainties are of the same magnitudo: 
$c\Delta_s t\sim \Delta_s {\eta}^i\sim \Delta_s {\eta}\sim \Delta_s L$. STUR in this case thus reduce to 
$\Delta_s L\geq \xi L_P$, where the constant $\xi$ is depending on the approximations made for the STUR and on the exact expression for the maximal localizing states and cannot be calculated in an exact way from the
semiclassical model in \cite{1}. From time-energy uncertainty translated in a Friedmann context we have
$\Delta_s E \Delta_s t\geq \frac{\hbar}{2}$. In a maximal localizing state with $\Delta_s L\geq \xi L_P$ we get
$\Delta_s E\sim\xi\frac{c\hbar}{2\Delta_s L}$. Hence, the generalized Misner-Sharp energy $E(L)$ at the physical scale $L$ can be written as the
classical expression summed with the spread due to quantum fluctuations given by $\xi\frac{c^4 L_P^2}{2G L}$, where $\Delta_s L\sim L$ and with
$\hbar=c^3L_P^2/G$.\\
As a consequence of the reasonings above,
after considering a given physical scale provided by an arbitrary spherical region of proper areal radius $L$, for the generalised Misner-Sharp energy $E(L)$, the effective energy density ${\rho}_{{\overline{\Lambda}}_L}$
and the effective cosmological constant ${\overline{\Lambda}}_L$ at the physical scale $L$ in our semiclassical model are given by
\begin{eqnarray}
& &E(L)=\frac{c^4}{2G}\frac{L^3}{L_{\Lambda}^2}+\xi\frac{c^4}{2G}\frac{L_P^2}{L},\;L_{\Lambda}=\sqrt{\frac{3}{\Lambda}},
\label{5}\\
& &{\rho}_{{\overline{\Lambda}}_L}=\frac{c^2\Lambda}{8\pi G}+\frac{3\xi c^2}{8\pi G}\frac{L_P^2}{L^4},
\label{6}\\
& &{\overline{\Lambda}}_L=\Lambda+\frac{3\xi L_P^2}{L^4},
\label{7}
\end{eqnarray}
where $E(L)$ in (\ref{5}) denotes the first member of equation (\ref{2}). 
Equation (\ref{6}) is obtaned from equation (\ref{5}) by divising for the volume $V$ of the 
spherical region of proper areal radius $L$, while equation (\ref{7}) is obtained multiplying (\ref{6}) for 
$8\pi G/c^2$. The parameter $\xi$ originates from spacetime uncertainty relations dictated by non-commutativity at Planckian scales (see \cite{1} for more details).
The effective cosmological constant (\ref{7}) is thus averaged on bigger and bigger scales. In practice, quantum fluctuations are very strong at Planckian scale $L_P$, but they become monotonically decreasing with increasing physical scale $L$. This is true up to the decoherence scale $L_D$, defined in \cite{1} as the scale where we have the crossover to classicality and fixed by the absolute minimum of (\ref{5}), given by $L_D={\left(\frac{\xi L_P^2}{\Lambda}\right)}^{\frac{1}{4}}$. Quantum modes are thus frozen at the physical scale $L_D$, representing the lowest energy scale
for the generalized  Misner-Sharp mass. Note that without the quantum-fluctuations motivated term in (\ref{5}), the classical Misner-Sharp energy is a monotonically
increasing function of $L$.
As a consequence, the observed value for $\overline{\Lambda}$ is fixed at
$L=L_D$ and is given by 
\begin{equation}
\overline{\Lambda}(L=L_D)=\overline{\Lambda}=4\xi\frac{L_P^2}{L_D^4}. 
\label{7b}
\end{equation}
The constant $\xi$ has been fixed 
in \cite{3} by 
analogy with Casimir effect to be $\xi=\frac{\pi^3}{360}$. It is important to stress that at any fixed scale $L$, the equation of state for the effective averaged ${\overline{\Lambda}}_L$ is exactly the one of the cosmological constant, i.e. 
$c^2{\rho}_{{\overline{\Lambda}}_L}=-p_{{\overline{\Lambda}}_L}$. According to general relativity, this does imply that at any fixed physical scale
$L$, the Newtonian constant $G$ does not depend on the cosmic time $t$. Also note that equation (\ref{7b}) fixes a relation between $\overline{\Lambda}$ and
$L_D$. Unfortunately, we have not a sound way to fix $L_D$: from (\ref{7b}) one cannot calculate $\overline{\Lambda}$ from $L_D$. Conversely, by knowing 
$\overline{\Lambda}$ we can infer the value of $L_D$. This is a limitation of our semiclassical approach: we could infer the value $L_D$ only within a complete
quantum gravity theory.\\
The existance of a decoherence scale is a fundamental ingredient of our model. This is a physically resonable concept. In practice, we assume
that quantum fluctuations are not working at cosmological scales, up to the whole visible universe, as supposed in many approaches addressing the cosmological constant issue (see for example \cite{15,16,17} and references therein). The modes propagating at scales greater than $L_D$ are frozen in the lowest energy configuration provided by the minimum of (\ref{5}), it denoting the crossover to classicality. The physical effect of quantum fluctuations in 
our model is, in some sense, watered down at scales greater than the Planck one. 
This procedure is, as an example, compatible with the usual Casimir effect where, in the case of a sphere of areal radius
$L$, the Casimir energy (see \cite{3} and
references therein) is given by $E_C=\hbar\frac{c\pi^3}{720 L}$ and thus it depends on the physical scale under consideration, in perfect analogy with the
equation (\ref{5}). 
Our approach is also reminescent of the usual renormalization group approach. In particular, thanks to equation (\ref{6})
naturally emerges a renormalization scale $\mu$ denoting the physical scale under consideration. In the following we perform this analogy in a cosmological curved background, where, as well known, a renormalization-group theory is not well established as in the flat case.

\section{Renormalization-group approach in a cosmological context}

In many approaches of the dark energy problem (see for example \cite{17,15,16,22}), the cosmological constant is supposed to vary with the cosmic 
time $t$. The price to pay, although experimental data at cosmological scales have not ruled out
this possibility, is that the Newtonian constant is time-varying, in disagreement with general relativity, or that
a coupling between dark matter and dark energy must be imposed. In these frameworks, the cosmological constant runs at any scale and cannot be easily fixed. There, the point is how to appropriately choose the renormalization group scale $\mu$. Generally, the cosmological constant gives contributions from zero-point energies of many quantum fields. As stated for example in \cite{17}, the cosmological constant does appear as a parameter in the Lagrangian and as a result it should be renormalized. In a renormalization group approach, the  quantity of interest is given by
\begin{equation}
\mu\frac{\partial{\rho}_{{\overline{\Lambda}}_\mu}}{\partial\mu}=\beta(\mu), 
\label{8}
\end{equation} 
where ${\rho}_{{\overline{\Lambda}}_\mu}$ denotes the renormalized cosmological constant at the scale $\mu$.
At sufficiently small\footnote{In \cite{17,15} this translates in the condition that $\mu$ is smaller than the mass
of the lightest quantum field. This condition will be specified in our model in the next section.} renormalization scales $\mu$, equation (\ref{8}) can be 
written \cite{17,16} in the following general way:
\begin{equation}
{\rho}_{{\overline{\Lambda}}_\mu}(\mu)={\rho}_{{\overline{\Lambda}}_\mu}(\mu=0)+A\mu^2+B\mu^4+o(1).
\label{9}
\end{equation}
The crucial point is the choice of the scale $\mu$. Some authors \cite{23} argued that could exist an infrared cutoff, defined in terms of a momentum scale $k$,
that stops at a given scale the running of $\Lambda$ in a renormalization group approach. Moreover, in the limit for $k\rightarrow 0$ an infrared attractive non-Gaussian fixed point 
could exist both for $\Lambda(k)$ and the renormalized Newtonian constant $G(k)$. It should be noticed that 
it is not a simple task to determine a physical mechanism underlying a given infrared cutoff. Moreover, there not exist in the literature a rigorous proof of the existence of
such a IR fixed point, in contrast with the UV fixewd point.
In any case, it is expected that $\mu$ denotes the typical graviton momenta present in a cosmological background \cite{15}. 
In the literature regarding the modeling of the cosmological constant, the mainstream of researchers (see for example 
\cite{17,17a,15,16,16a,22,23}) assumes a time evolving renormalization scale $\mu$. In \cite{17,17a,16,16a} it 
has been assumed for the parameter $\mu$ that $\mu\sim H(t)$, i.e. varying with the Hubble flow, while in \cite{15} the identification 
$\mu\sim{\rho}^{\frac{1}{4}}(t)$ has been investigated. In particular, the choice in \cite{15} does imply that the renormalization scale parameter $\mu$ cannot be less than the one representing the typical momenta of background gravitons. Moreover,
it is rather questionable that quantum effects can survive at cosmological scales. It seems more realistic that 
quantum effects can be extended up to a certain scale $\mu_D$ at which the crossover to classicality emerges.\\
At this point it is important to note that we have depicted a model for $\overline{\Lambda}$ in a de Sitter background where the choice $\mu=const$ is rather natural since
$H_{\overline{\Lambda}}=const$. However, also in a more general FLRW spacetime equipped with matter and radiation with density respectively denoted with
${\rho}_m$ and ${\rho}_{rad}$ the term $\overline{\Lambda}$ continues, at a classical level, to be constant both in space and time. 
If we are wilings to accept that in a de Sitter universe $\overline{\Lambda}$ is a constant, then we are also legitimate to suppose that in a more general 
FLRW spacetime $\overline{\Lambda}$ remains constant both in space and time, with the actually measured constant value determined at the scale $L_D$. 
With respect to this point, we can justify our setup  with the identification $\mu\sim\frac{1}{L}=const$ also in a FLRW background. In a flat FLRW spacetime we have
\begin{equation}
H^2=\frac{8\pi G}{3c^2}\left({\rho}_m+{\rho}_{rad}\right)+\frac{\overline{\Lambda}c^2}{3},\;\;E_{ms}=\frac{c^2}{2G}L^3 H^2.
\label{k1}
\end{equation}
As a consequence of (\ref{k1}), for the generalized Misner-Sharp energy we have:
\begin{equation}
E_{ms}=\left({\rho}_m+{\rho}_{rad}\right)\frac{4\pi}{3}L^3+\frac{c^4}{6G}L^3{\overline{\Lambda}}_L,
\label{k2}
\end{equation}
with ${\overline{\Lambda}}_L$ given by (\ref{7}). 
If we suppose that, at least at a semiclassical level, renormalization due to quantum Planckian fluctuations mainly affects the cosmological constant, then
we can assume that the scale $L_D$ is determined by the absolute minimum with respect to $L$ of the terms in (\ref{k2}) of $E_{ms}$ independent on 
${\rho}_m$ and ${\rho}_{rad}$ and representing the vacuum energy part of Misner-Sharp energy, i.e.
provided by the minimum of the term
$\frac{c^4}{6G}L^3{\overline{\Lambda}}_L$ also in a FLRW context, thus leaving the expression (\ref{7b}) unchanged. We stress that this reasoning is 
justified by the physical fact that the present day concordance cosmological model at macroscopic scales is well depicted in terms of matter-energy density living together with 
a strictly constant comological constant. As a result, although the choice for a running cosmological constant is the one most used in the literature,
it is also physically reasonable to assume a renormalization procedure dressing the cosmological constant but leaving this a true constant.\\
As we see in the next section, we will use the identification $\mu\sim\frac{1}{L}$ to justify
the presence of $L_D$. To the best of my knowledge, this choice is new in the literature, at least in a cosmological context
\footnote{Note that in \cite{23a} a similar identification has been supposed, but at the physical scale represented by galaxies}, 
but completely natural in our model, where an effective cosmological constant ${\overline{\Lambda}}_L$ given by (\ref{7}) is introduced at the physical scale $L$.

\section{Renormalization scale in our model and its physical meaning}

As noticed in section above, the parameter $\mu$ denotes the renormalization scale or subtraction point where the physics is calculated and 'fixed'. In our 
semiclassical model \cite{1,2,3}, an effective model is obtained by adapting \cite{2} the fitting problem of modern cosmology to microscopic scales. The generalized
Misner-Sharp energy (\ref{5}) with energy density (\ref{6}) and the cosmological constant (\ref{7}) are thus effective quantities obtained solving the fitting problem at 
a given fixed microscopic scale $L$. It is thus natural to suppose a certain relation between $\mu$ and $L$, i.e. $\mu=\mu(L)$. To be more explicit, we must compare equations 
(\ref{6}) and (\ref{9}). Hence, in order to depict a cosmological constant strictly constant and not time dependent, we are forced to pose $\mu=\frac{L_P}{L^2}$. 
Note that $\mu$ has the dimension of an inverse of a length and as a result the quantity $\hbar\mu$
has the dimension of a momentum. It is also expected that the quantity $\hbar\mu$ is a measure of typical graviton momenta. As a result of this choice the (\ref{6}) 
becomes
\begin{equation}
{\rho}_{{\overline{\Lambda}}_{\mu}}=\frac{c^2\Lambda}{8\pi G}+\frac{3\xi c^2}{8\pi G}\mu^2.
\label{10}
\end{equation}
By a comparison with (\ref{9}) we can write 
\begin{equation}
{\rho}_{{\overline{\Lambda}}_{\mu}}(\mu=0)=\frac{c^2\Lambda}{8\pi G},\;\;\;\;\;\;A=\frac{3\xi c^2}{8\pi G}.
\label{11}
\end{equation}
Expression (\ref{9}) suggests that the expression (\ref{6}) for ${\rho}_{{\overline{\Lambda}}_L}$ represents the principal part of a more general expression obtained by considering higher order terms in a series expansion
with respect to $\mu=\frac{L_P}{L^2}$. 
By inspection of (\ref{9}) we deduce that (\ref{6}) can be generalized, according to \cite{17}, in the following way:
\begin{equation}
{\rho}_{{\overline{\Lambda}}_{\mu}}=\frac{c^2\Lambda}{8\pi G}+\frac{3\xi c^2}{8\pi G}\mu^2+\frac{3\xi_2 c^2}{8\pi G}\mu^4+\cdots+\frac{3\xi_n c^2}{8\pi G}\mu^{2n}
+o(1),
\label{12}
\end{equation}
where $\{\xi_n\}$ are real parameters and $\Lambda$ denotes the bare cosmological constant.
Note that, according for example with \cite{17,17a,16,16a,15}, only even powers of $\mu$ are present in the series expansion.
The expansion (\ref{12}) is in terms of the above chosen parameter $\mu$. As we see later in the paper,
the higher order terms in (\ref{12}) can be interpreted as counterterms added to
(\ref{6}) and due to the renormalization procedure: these terms dress the cosmological constant at the IR fixed point $\mu_D$ to its observed value
$\overline{\Lambda}$. 
A necessary condition for the convergence of this
expansion series is achieved when $L>a\simeq{10}^{-18}\;meters$  ($\mu<1$). 
This estimation simply means that the necessary condition for the convergence of expansion (\ref{12}) is certainly valid near the decoherence 
scale $L_D$ that, according to \cite{1,2,3} is about ${10}^{-5}\;meters$. More generally, by supposing the Planckian ultraviolet cutoff $L_P$ for a minimal length due to the 
possible non-commutative nature of the spacetime at Planckian scales, and the existence of a decoherence scale we have:
\begin{equation}
\mu\in \left[\frac{L_P}{L_D^2},\frac{1}{L_P}\right].
\label{14}
\end{equation}
According to the physical meaning of $\mu$, equation (\ref{14}) does imply that $\mu_D=\frac{L_P}{L_D^2}$ is the minimum allowed value for the renormalization-group 
scale parameter $\mu$. Physically, quantum modes propagating with momentum less than $\hbar\mu_D$ should be truncated: these modes are thus 'frozen' it denoting
the crossover to classicality.\\
In terms of the $\beta$ function, from (\ref{12}) we have:
\begin{equation}
\mu\frac{d{\rho}_{{\overline{\Lambda}}_{\mu}}}{d\mu}=\beta(\mu)=\frac{3\xi_1 c^2}{4\pi G}\mu^2+o(1).
\label{15}
\end{equation}
From (\ref{15}) we deduce that $\beta(\mu)=0$ for $\mu=0$. Thus we formally should have a trivial IR fixed point obtained at $\frac{L_P}{L^2}\rightarrow0$: 
as a result this value for $\mu$ can be 
strictly obtained in the limit for $L\rightarrow\infty$. Consequently, one should expect that the extrapolated value of the cosmological constant approaches asymptotically to zero.
However, as supposed in \cite{1,2,3}, we do not assume 
that quantum fluctuations can act on macroscopic distances up to the whole visible universe.
Rather, we assume that quantum modes propagating above certain physical scales or below a certain momentum are truncated: this is what 
identifies the decoherence scale $L_D$. This scale acts, in our semiclassical model, as a kind of effective non-trivial fixed point where the observed value of the 
cosmological constant is renormalized. However, it should be stressed that our purpose is not to rigorously show the existence of an IR fixed points, but rather
that the possible existenxe of an IR fixed point supports the existence of $L_D$. 
To further support this statement, consider a de Sitter universe (\ref{1}) thus equipped with the observed cosmological constant $\overline{\Lambda}$ and Hubble
(apparent\footnote{This is the marginally outer trapped surface.}) horizon $L_{\overline{\Lambda}}$ given by 
$L_{\overline{\Lambda}}=\sqrt{\frac{3}{\overline{\Lambda}}}=\frac{c}{H_{\overline{\Lambda}}}$. Moreover, we consider for simplicity the expression (\ref{10}), thus neglecting
higher order terms with respect to $\mu$.
As a consequence, thanks to equation $\overline{\Lambda}(L=L_D)=\overline{\Lambda}=4\xi\frac{L_P^2}{L_D^4}$ with
$\xi=\frac{\pi^3}{360}$, for the renormalization scale parameter at the decoherence scale $\mu_D$ we obtain
\begin{equation}
\mu_D=\frac{L_P}{L_D^2}=\frac{30}{\pi^3}\frac{H_{\overline{\Lambda}}}{c}\simeq\frac{H_{\overline{\Lambda}}}{c}=\frac{1}{L_{\overline{\Lambda}}}.
\label{16}
\end{equation} 
The result (\ref{16}) can be interpreted in light of the reasonings above. After multiplying $\mu$ for
$\hbar$, the quantity $\hbar\mu$ denotes the typical momentum of gravitons present in the background: we have 
$\hbar\mu\in \left[\frac{\hbar L_P}{L_D^2},\frac{\hbar}{L_P}\right]$. Hence, we obtain the result that the minimum allowed gravitons momentun 
$P_{g_{min}}$ is provided by $P_{g_{min}}\simeq\frac{\hbar}{L_{\overline{\Lambda}}}$ that in turn is the one provided by gravitons with the maximum allowed
wavelength corresponding to the proper radius of the apparent horizon in a de Sitter universe.\\ 
At first look, since the scale of $\mu_D$ is a cosmological scale,
it seems that the arguments leading to (\ref{16}) is a simple way to obtain a cosmological scale from our semiclassical model, but this is 
not the case. In fact, the relation (\ref{7b}) is dictated by the minimum of (\ref{5}) with respect to $L$, while the choice of $\mu$ is forced by
the close analogy between (\ref{6}) and (\ref{9}). The term proportional to $1/L$ in (\ref{5}) is motivated by quantum gravity arguments at Planckian scales
(see \cite{1}) and thus it is non correlated from the onset to macroscopic scales. To this purpose, note that the scale
$L_D$ is the one where the two terms in (\ref{5}) have the same magnitudo. Also in the usual Casimir effect, the scale $10^{-5}$ meters 
(see \cite{3} and references therein) is the one where the quantum
fluctuations have the same magnitudo of the statistical ones: hence above this scale quantum fluctuations are dominated by statistical fluctuations.\\
On general grounds, it seems rather reasonable to suppose that a complete 
quantum gravity theory could predict a $\beta$ function for ${\rho}_{{\overline{\Lambda}}_{\mu}}$ vanishing at $\mu=\mu_D$. To this purpose note that the parameter
$\mu$ at the decoherence scale is rather small: $\mu_D\simeq 10^{-26}/meters$. It is also reasonable to suppose that with a suitable modification of 
the series expansion (\ref{12}), we can formally write the expression for the 'renormalized' ${\rho}_{{\overline{\Lambda}}_{\mu}}$ in a neighborhood of the 
decoherence scale $\mu_D$ in term of a series expansion near $\mu=\mu_D$ in the form:
\begin{equation}
{\rho}_{{\overline{\Lambda}}_{\mu}}=\frac{c^2\overline{\Lambda}}{8\pi G}+A_2{\left(\mu-\mu_D\right)}^2+A_4{\left(\mu-\mu_D\right)}^4+o(1).
\label{17}
\end{equation}
Obviously, a rigorous proof for (\ref{17}) can be only possible within explicit Quantum Field Theory calculations that is outside the purposes of the
present paper.
How can we physically justify the expression  (\ref{12}) or (\ref{17}) ?
In our semiclassical model expressions (\ref{5}) and (\ref{6}) are related by the relation $E(L)={\rho}_{{\overline{\Lambda}}_{L}}V(L)$ with
$V(L)=\frac{4\pi}{3}L^3$. To this purpose, note that the generalized Misner-Sharp quasi-local energy expression $E(L)$ in (\ref{5}) has been obtained in
\cite{1} by considering a semiclassical model involving quantum fluctuations dictated by quantum-motivated reasonings at 
the physical scale $L$. Hence, formula (\ref{5}) is a way to 'dress'
the classical Misner-Sharp energy by taking into account quantum fluctuations by the term proportional to $1/L$. As a first approximation it is reasonable to suppose that the relation $E(L)={\rho}_{{\overline{\Lambda}}_{L}}V(L)$ still holds, at least on average: this means that the so obtained 
energy density ${\rho}_{{\overline{\Lambda}}_{L}}$ can be considered as a mean field approximation on a volume of proper areal radius $L$.
However, it is also expected that corrections to the relation $E(L)={\rho}_{{\overline{\Lambda}}_{L}}V(L)$ arise due to higher order correction
terms. Consequently, with the expression
(\ref{5}) for $E(L)$, we expect:
\begin{equation}
{\rho}_{{\overline{\Lambda}}_{L}}=\frac{E(L)}{V(L)}+corrections.
\label{r1}
\end{equation}
The corrections in (\ref{r1}) are exactly given, according to (\ref{9}) and thanks to (\ref{12}), by 
\begin{equation}
corrections=\frac{3\xi_2 c^2}{8\pi G}\frac{L_P^4}{L^8}+\cdots+\frac{3\xi_n c^2}{8\pi G}\frac{L_P^{2n}}{L^{4n}}+\cdots.
\label{r2}
\end{equation}
The expression 
(\ref{17}) is the exptected form, thanks to (\ref{9}), for the renormalization group expression for the energy density in the infrared limit, i.e.
the (\ref{17}) it gives the renormalized energy density (and thus the renormalized cosmological constant) 
in a neighborhood of decoherence scale, where $\mu_D$ 
is in turn a fixed point for the beta function $\beta(\mu)$:
\begin{equation}
\beta(\mu)=2A_2\mu(\mu-\mu_D)+o(1).
\label{18}
\end{equation}
The idea is thus to add, in the spirit of the renormalization 
group approach, counterterms in (\ref{6}) dressing the 
cosmological constant and thus rewrite the expression (\ref{12}) so obtained in a suitable way, namely equation (\ref{17}).\\ 
First of all, note that by using the Newton's binomial, we have:
\begin{equation}
{\left(\mu-\mu_D\right)}^n=\sum_{k=0}^{n}\frac{n!}{k!(n-k)!}{\mu}^{n-k}{(-1)}^k{\mu}_D^{k}.
\label{19}
\end{equation}
In order to modify (\ref{12}), note that expression (\ref{19}) contains a term linear in $\mu$, terms with odd powers in
$\mu$ and a term depending only on the constant $\mu_D$ obtained with $n=k$. Since to obtain the (\ref{17}) a summation of the (\ref{19}) with respect to
$n$ must be performed, we expect a complicated expression where any power $\mu^m$ in (\ref{17}) takes contributions $\forall n\geq m$. The term independent  of $\mu$ 
can be obtained from (\ref{12}) with the transformation:
\begin{equation}
\Lambda\rightarrow \overline{\Lambda}+\sum_{i=1}^{\infty}k_i{\mu}_D^{2i},\;\;k_i\in\Re,
\label{20} 
\end{equation}
where the constants $\{k_i\}$ take obviously contributions from any term of the series expansion (\ref{17}). Obviously, at $\mu=\mu_D$, only the renormalized term 
$\frac{c^2\overline{\Lambda}}{8\pi G}$ in (\ref{17}) survives.
Terms with odd powers in $\mu$ could be obtained at hand from the 'bare'
energy density by adding all terms with odd powers in $\frac{L_P}{L^2}$, but are in disagreement with (\ref{9}) 
(see for example \cite{17,17a,16,16a} and references therein): 
as a consequence we must found a procedure to eliminate such odd terms. 
Finally, note that a linear term  $\mu$ is present in (\ref{17}). Unfortunately, this term is not explicitely present in (\ref{12}). 
In the next section we outline a procedure to eliminate odd powers in expansion (\ref{12}).

\section{Building a non-trivial IR fixed point at $\mu=\mu_D$}

In this section we formally derive the expression for the $\beta$ function in order to have a non trivial IR fixed point and thus does not represent a rigorous proof
of the existence od such a IR fixed point.\\
To start with, it is more convenient to work with adimensional constants: to this purpose we define
$A_{2n}=\frac{c^2}{8\pi G}\frac{D_{2n}}{{\mu}_D^{2n-2}}$. The (\ref{17}) becomes
\begin{equation}
{\rho}_{{\overline{\Lambda}}_{\mu}}=\frac{c^2\overline{\Lambda}}{8\pi G}+
\frac{c^2}{8\pi G}\left[\sum_{n=1}^{\infty}\frac{D_{2n}}{{\mu}_D^{2n-2}}{\left(\mu-\mu_D\right)}^{2n}\right].
\label{21}
\end{equation}
The series expansion (\ref{21}) is valid in a sufficiently small neighborhood of the decoherence scale $\mu_D$. Note that the (\ref{21}) takes formally evident the existence of
a non trivial fixed point at $\mu=\mu_D$, while in the form (\ref{12}) only the trivial fixed point at $\mu=0$ is evident.
The strategy is to take the series expansion (\ref{21}) and thus compare it with the expansion (\ref{12}). The conditions 
are obviously the (\ref{20}), the ones concerning
the vanishing of the odd powers in $\mu$ and finally the one involving the value of $\xi$ that has been fixed in \cite{3} to be $\xi=\frac{\pi^3}{360}$. By formula
(\ref{19}), we deduce the following general structure for the linear system above defined:
\begin{eqnarray}
& &N_2^0 D_2+N_4^0 D_4+\cdots+N_{2n}^0 D_{2n}+\cdots=3\xi\label{22}\\
& &N_2^1 D_2+N_4^1 D_4+\cdots+N_{2n}^1 D_{2n}+\cdots=0 \nonumber\\
& &N_4^2 D_4+N_6^2 D_6+\cdots+N_{2n}^2 D_{2n}+\cdots=0 \nonumber\\
& &N_6^3 D_6+N_8^3 D_8+\cdots+N_{2n}^3 D_{2n}+\cdots=0 \nonumber\\
& &\vdots\;\;\;\;\;\;\;\;\;\;\;\;\;\;\;\;\;\;\;\;\;\;\;\;\;
\;\;\;\;\;\;\;\;\;\;\;\;\;\;\;\;\;\;\;\;\;\;\;\;\;\;\;\;\;\;\vdots=0\nonumber\\
& &N_{2k}^m D_{2k}+\cdots+N_{2n}^m D_{2n}+\cdots=0\nonumber\\
& &\vdots\;\;\;\;\;\;\;\;\;\;\;\;\;\;\;\;\;\;\;\;\;\;\;\;\;
\;\;\;\;\;\;\;\;\;\;\;\;\;\;\;\;\;\;\;\;\;\;\;\;\;\;\;\;\;\;\vdots=0,\nonumber
\end{eqnarray}
where the coefficients $\{N_{2k}^m\}$ are determined by a suitable summation of binomial coefficients of (\ref{19}). 
The second equation of (\ref{22}) cancels out the term proportional to $\mu$, while the third equation cancels out the term proportional to
$\mu^3$ and so on.\\
From the structure of the system (\ref{22}), it results that 
we could obtain, from the first of (\ref{22}) the constant $D_2$ in terms of $\xi$ and of the other variables $\{D_{2n}\}$ as an infinite series and subsitute the expression so obtained 
in the second equation of the system. The procedure continues by formally applying the well known Gauss procedure. As a result, a very complicated iterative expression arises that 
cannot be recast in an explicit analytic formula. Fotunately, the system (\ref{22}) can be solved by fixing a given order in the series expansion (\ref{21}).\\
To show this procedure, we firstly consider the series expansion (\ref{21}) with two terms: 
\begin{equation}
{\rho}_{{\overline{\Lambda}}_{\mu}}=\frac{c^2\overline{\Lambda}}{8\pi G}+
\frac{c^2}{8\pi G}\left[D_2{\left(\mu-\mu_D\right)}^{2}+\frac{D_4}{{\mu}_D^2}{\left(\mu-\mu_D\right)}^{4}\right]+o(1).
\label{23}
\end{equation}
The terms in square bracket of (\ref{23}) become:
\begin{eqnarray} 
& &(D_2+D_4)\mu_D^2+\mu\left(-2D_2-4D_4\right)\mu_D+\label{24}\\
& &+\mu^2\left(D_2+6D_4\right)-4\mu^3\frac{D_4}{\mu_D}+\mu^4\frac{D_4}{\mu_D^2}.\nonumber
\end{eqnarray}
From (\ref{24}) it is evident that we can cancel out the term linear in $\mu$ but not the one proportional to $\mu^3$ since this term depends only on
the constant $D_4$. Consequently, with the two terms in (\ref{23}) only the expansion up to $\mu^2$ is obtainable. To cancel out the term proportional to
$\mu^3$ we must consider also the term with $n=3$ in (\ref{21}). 
Hence, system (\ref{22}) with (\ref{23}) and (\ref{24}) becomes:
\begin{eqnarray}
& &D_2+6D_4=3\xi\label{25}\\
& &-2D_2-4D_4=0,\nonumber
\end{eqnarray} 
with solution $D_2=-\frac{3}{2}\xi,\;D_4=\frac{3}{4}\xi$. Up to the order $\mu^2$ we obtain:
\begin{equation}
{\rho}_{{\overline{\Lambda}}_{\mu}}=\frac{c^2}{8\pi G}\left(\overline{\Lambda}-\frac{3}{4}\xi\mu_D^2\right)+\frac{3\xi c^2}{8\pi G}\mu^2+o(1).
\label{26}
\end{equation}
Equation (\ref{26}) must be compared with (\ref{12}), with the (\ref{20}) given by 
\begin{equation}
\Lambda\rightarrow \overline{\Lambda}-\frac{3}{4}\xi\mu_D^2.
\label{27}
\end{equation}
In a similar manner, to obtain the term proportional to $\mu^4$ we must consider a further term in (\ref{21}):
\begin{equation}
{\rho}_{{\overline{\Lambda}}_{\mu}}=\frac{c^2\overline{\Lambda}}{8\pi G}+
\frac{c^2}{8\pi G}\left[\sum_{n=1}^{3}\frac{D_{2n}}{{\mu}_D^{2n-2}}{\left(\mu-\mu_D\right)}^{2n}\right]+o(1).
\label{28}
\end{equation}
In this case system (\ref{22}) gives:
\begin{eqnarray}
& & D_2+6D_4+15D_6=3\xi\label{29}\\
& &-2D_2-4D_4-6D_6=0\nonumber\\
& &-4D_4-20D_6=0,\nonumber
\end{eqnarray}
with solution $D_2=-\frac{21}{8}\xi,\;D_4=\frac{15}{8}\xi,\;D_6=-\frac{3}{8}\xi$. We can thus write:
\begin{equation}
{\rho}_{{\overline{\Lambda}}_{\mu}}=\frac{c^2}{8\pi G}\left(\overline{\Lambda}-\frac{9}{8}\xi\mu_D^2\right)+\frac{3\xi c^2}{8\pi G}\mu^2
-\frac{\xi c^2}{32\pi G}\frac{15}{\mu_D^2}\mu^4+o(1).
\label{30}
\end{equation}
The two examples above are enough to illustrate the method. It should be noticed, obviously, that with the tecnique depicted above, we can obtain the 
fixed point at $\mu=\mu_D$ only asymptotically, for $n\rightarrow\infty$ by summing up all terms in the series (\ref{21}). In fact, with (\ref{26}) we formally obtain the
trivial fixed point at $\mu=0$, while with the (\ref{30}) we again formally obtain the trivial fixed point and a second one with 
$\mu=\sqrt{\frac{2}{5}}\mu_D$, asymptotically approaching the fixed point $\mu=\mu_D$, i.e. in the limit for $n\rightarrow\infty$. 
A necessary condition for the convergence of (\ref{21}) is  $\mu\in(0,2\mu_D)$. Also note that parameter $\mu=\mu_D$ in (\ref{21}) and thus in (\ref{26}) and
(\ref{30}) is the one obtained by searching an absolute minimum for (\ref{12}). Moreover, the coefficients of all powers $\mu^{2n}$,
exception made\footnote{This is because the coeffcient $\xi$ has been fixed in \cite{3}.} for the term $\mu^2$, are expressed as a
series expansion and thus take more contributions order by order, i.e. only in the limit for $n\rightarrow\infty$ we have the exact coefficients.
Also note that, thanks to (\ref{19}), the first equation of system (\ref{22}) becomes:
\begin{equation}
\sum_{n=1}^{\infty}n(2n-1)D_{2n}=3\xi.
\label{30a}
\end{equation}
Equality (\ref{30a}) does imply that series in the left side is convergent and as a consequence we must have $D_{2n}=o(1/n^2)$ for $n\rightarrow\infty$: this
fact garanties the convergence of the series expansion (\ref{21}) sufficiently near $\mu_D$.\\
The reasonings above show that our procedure works. 
It is interesting to note that the condition to formally have a non trivial fixed point forces the constant $\{\xi_n\}$ in (\ref{12}) to be proportional to the
'phenomenological' constant $\xi$ that has been fixed in \cite{3} by a comparison with the Casimir effect with  a spherical cavity.
Also note that the $\beta$ function so obtained, thanks to the series expansion (\ref{21})),
is strictly positive approaching $\mu_D$. Hence, the fixed point at $\mu=\mu_D$ is an IR one, with the coupling constant
$\overline{\Lambda}$ decreasing with decreasing energy, i.e. we have a non-trivial IR-stable fixed point.\\
 
\section{Conclusions and final remarks}
In this paper we continued the investigations in \cite{1,2,3} concerning a proposal on the nature of the cosmological constant. 
In particular, this paper is focused on
a possible relation between our formulas regarding the energy density and the ones outlined in the literature, namely equations (\ref{8}) and
(\ref{9}), as an attempt to reproduce a renormalization group approach for $\overline{\Lambda}$ in a cosmological 
background. To this purpose, the crucial point is
the choice of the renormalization scale parameter $\mu$. In the literature (see for example \cite{17,17a,16,16a,15,23} and references therein) the parameter $\mu$ 
typically is chosen as a function of the Hubble flow $H$ \cite{17,16} or in terms of the cosmic temperature $T\sim\sqrt{\rho}^{\frac{1}{4}}$. 
The price to pay in 
these cases is a time dependent cosmological constant. In the literature \cite{23b}
we can find, for example, studies where a time evolving $\Lambda$ is coupled with dark energy.
Moreover, a time evolving $\Lambda$ is often related to the variability of $G$, as for example in Brans-Dicke theories.\\
In our model the parameter $\mu$, thanks to (\ref{6}) and (\ref{9}), is chosen to be $\mu=\frac{L_P}{L^2}$. Moreover, equations (\ref{12}) formally does imply a 
trivial fixed point at $\mu=0$, i.e $L\rightarrow\infty$. A trivial IR  fixed point is expected, as an example, for free gravitons traveling in a 
Minkowskian spacetime. 
If the decoherence scale $L_D$ denoting the crossover to classicality is really existing, 
then it is expected that in a renormalized theory an IR-stable fixed point for $\overline{\Lambda}$ should arise for $\mu=\mu_D$. In section 5, we have shown that this task can 
be formally accomplished within our approach, provided that  corrections
to the expression (\ref{6}) are considered. Consequently, in order to relate the series expansion (\ref{21}) with (\ref{12}),
a strategy has been outlined in section 5. To this regard, formula (\ref{20}) can be interpreted as the relation between the bare cosmological constant $\Lambda$ 
and the renormalized one $\overline{\Lambda}$ at the decoherence scale $L_D$. Also note that the series expansion (\ref{21}) for $\mu=0$ 
formally becomes:
\begin{equation}
{\rho}_{{\overline{\Lambda}}_{\mu=0}}=\frac{c^2\overline{\Lambda}}{8\pi G}+
\frac{c^2\mu_D^2}{8\pi G}\left[\sum_{n=1}^{\infty}D_{2n}\right].
\label{31}
\end{equation}
By a comparison with (\ref{12}) we deduce that:
\begin{equation}
\Lambda\rightarrow \overline{\Lambda}+\mu_D^2\left[\sum_{n=1}^{\infty}D_{2n}\right].
\label{32}
\end{equation}
As specified in \cite{1,2,3}, the quantity $\Lambda$ denotes the bare cosmological constant, while the observed one $\overline{\Lambda}$ is dressed by vacuum 
fluctuations and equation (\ref{32}) depicts the relation between  $\Lambda$ and $\overline{\Lambda}$.
Also for formula (\ref{5}), at the renormalization scale $\mu=\mu_D$,
we have:
\begin{equation}
 E(L_D,\Lambda)=\frac{c^4}{6G}{L_D^3 \Lambda}+\xi\frac{c^4}{2G}\frac{L_P^2}{L_D}\rightarrow E(L_D,\overline{\Lambda})=\frac{c^4}{6G}{L_D^3 \overline{\Lambda}},
\label{33}
\end{equation} 
where the expression $E(L_D,\overline{\Lambda})=\frac{c^4}{6G}{L_D^3 \overline{\Lambda}}$ is the classical one expected for $L\geq L_D$, 
i.e. the classical Misner-Sharp energy $E(L)=\frac{c^4}{6G}L^3 \overline{\Lambda}$.\\
As commented in section 4, before equation (\ref{17}), the scale $L=L_D$ is the scale such that modes propagating with $L>L_D$ are frozen in the lowest 
energy state for $E(L)$ and as a consequence classicality emerges.\\
Concerning the physical scale, in our model we used the proper length $L$. This is because in a general relativistic context proper lengths are 
appropriate quantities to describe lengths. When we consider a given fixed physical scale, we mean $L=cosnt$ and as a result the proper areal radius 
is held fixed in time. In a de Sitter background (\ref{1}) this does imply $L=a(t)R=const$, where $R$ is the comoving radius. Hence $R$ is time varying but
$a(t)R(t)$ is fixed to the desired scale. As an example, it is obvious the meaning of fixing the Planck scale $L_P$ in a Minkowskian spacetime, while the 
meaning of $L_P$ in a curved spacetime must be specified. In our setup we have $L_P\simeq 10^{-35}\;meters=a(t)R(t)$. As a consequence, the comoving
distance for the Planck length is era-dependent but not the proper Planck length $L_P$.\\ 
As a final but necessay consideration, in relation to the existence of a decoherence scale $L_D$, it should be noticed that gravity has been tested up to distances of  order of $10^{-2}$ meters \cite{24} and consequently a decoherence scale fixing the observed value of $\overline{\Lambda}$
of the order of $10^{-5}$ meters, as estimated in \cite{1}, can be reasonable. From a physical point of view, it is possible to consider spacetimes fluctuations dictated by quantum gravity acting in a non-negligible way at Planckian scales. 
If we consider a given physical scale $L_D$, one can suppose that it is the quantized spacetime that decoheres, where the environment of a spherical region 
$L$ could be provided by a thermal bath of gravitons with a mean momentum of the order of $\hbar\mu_D$. To this purpose, note that in \cite{26} a statistical description
of $\overline{\Lambda}$ in a de Sitter universe has been proposed, where spacetime is filled with a bath of gravitons near a Bose-Einstein condensate with a mean
frequency $\overline{\omega}$ given by $\overline{\omega}\simeq 10^{-18} Hz$. It is a matter of fact that $\mu_D\simeq\overline{\omega}/c$. 
This picture, without a sound quantum gravity theory, is very difficult to implement in a more rigorous manner and a general tretment concerning decoherence caused by 
spacetime 
fluctuations is still lacking in the literature. For a nice review regarding decoherence in a gravitational context the reader can see \cite{25}
and references therein.\\ 
Summarizing, the presence of the invariant scale $\mu_D$ is a new
physically viable manner to justify a small cosmological constant $\overline{\Lambda}$ filling the whole visible universe with constant density both in space and time.

\end{document}